\documentstyle[12pt,epsf,aps]{revtex}  

\begin{document}
\newcommand{\p}{\partial}
\newcommand{\ls}{\left(}
\newcommand{\rs}{\right)}
\newcommand{\beq}{\begin{equation}}
\newcommand{\eeq}{\end{equation}}
\newcommand{\beqa}{\begin{eqnarray}}
\newcommand{\eeqa}{\end{eqnarray}}
\newcommand{\bdm}{\begin{displaymath}}
\newcommand{\edm}{\end{displaymath}}
\begin{center}
{\large {\bf
Radial flow of kaon mesons in heavy ion reactions
}}
\vskip 0.5 true cm
\vskip 0.5 true cm
Z.S.Wang, Amand Faessler, C. Fuchs, V.S.Uma Maheswari and T.Waindzoch
\vskip 0.5 true cm
\vskip 0.5 true cm
Institut f\"ur Theoretische Physik der Universit\"at T\"ubingen,\\
Auf der Morgenstelle 14, D-72076 T\"ubingen, Germany
\vskip 0.5 true cm
\vskip 0.5 true cm
\vskip 0.5 true cm
\centerline{\large\bf Abstract}
\vskip 0.5 true cm
\vskip 0.5 true cm
\end{center}
This work investigates the collective motion of kaons
in heavy ion reactions at SIS energies
(about 1-2 GeV/nucleon). A radial collective flow of $K^+$ mesons
is predicted to exist in central Au + Au collisions, which manifests
in a characteristic "shoulder-arm" shape of the
transverse mass spectrum of the midrapidity $K^+$ mesons.
The $K^+$ radial flow
arises from the repulsive $K^+$ mean field in nuclear matter.
In spite of a strong reabsorption and rescattering
the attractive $K^-$ mean field
leads as well to a collective radial flow of $K^-$ mesons.
The $K^-$ radial flow, however, is different from that of $K^+$ mesons
and can be observed by a characteristic "concave"
structure of the
transverse mass spectrum of the $K^-$ mesons emitted at midrapidity.
The kaon radial flows
can therefore serve as a novel tool for the investigation of kaon
properties in dense nuclear matter.
\vspace{5mm}\
\vskip 0.5 true cm
\vskip 0.5 true cm
25.75.+r
\newpage
\section{Introduction}

Collective motion in heavy ion reactions
has attracted increasing experimental and theoretical interests.
"Bounce-off"\cite{stoecker80,gustafsson84,danie85}, "squeeze-out"
\cite{stoecker82,gutb892} as well as radial flow\cite{lisa95,bear97,schne92}
of nuclear matter is predicted by theories and observed in experiments.
By "bounce-off" one means a deflection of the spectators sidewards
in the reaction plane, while "squeeze-out" denotes nuclear matter
jets in the direction perpendicular to the reaction plane emitted from the
high density overlap region ( so-called fireball). Radial flow is a
collective expansion of the compressed nuclear matter. 
The collective motion has proved to be a good probe
for the equation of state of nuclear matter (EOS)\cite{pan93,bass95}
as well as for the heavy ion reaction dynamics\cite{sorge97,volo97,barr94}.
\vskip 0.5 true cm
The techniques developed in studying the collective flow of nuclear matter
have also been applied to secondary particles such
as pions, kaons, hyperons etc.\cite{bli91,bass932,kint97,li95,wang97,ritman95,brill93,venema93,bass94,senger96,jahns94,kahana95,li96}.
For secondary particles
"squeeze-out" and "transverse flow" are defined in the same way as for
nucleons by the
azimuthal anisotropy and the rapidity
dependence of the in-plane component of the particle momentum, respectively.
The prediction of a pion transverse flow anticorrelated with
that of nucleons has been confirmed by experiments\cite{bli91,bass932,kint97}.
Kaon transverse flow attracts also attention
since it might provide
information about possible modifications of kaon properties in dense nuclear
matter\cite{li95,wang97,ritman95}.
The in-medium kaon properties
are an important issue not only for nuclear physics, but also for astrophysics
with respect to the evolution of compact stars\cite{brown94}.
Secondary particles are usually created from elementary
hadron-hadron collisions which take place randomly in the colliding system.
In the spirit of a proposal by
Fermi \cite{fermi50}, multiple particle productions from energetic
hadron-hadron collisions can be understood in terms of
a thermal description. Due to the fairly strong interaction between
particles involved
in a collision, the available phase space will be occupied according
to statistical laws. This physical picture has been supported
by extensive experiments on particle production from heavy ion reactions.
Therefore, a collective
motion of secondary particles should be identified by observing
deviations from the pure thermal
description.
It still remains an open question if secondary particles
can exhibit collective motion in heavy ion reactions in this strict
sense.

\vskip 0.5 true cm
In this paper, we explore the possible existence of a collective motion
of $K^+$ and $K^-$ mesons
in heavy ion reactions by comparing to a pure thermal picture.
We mainly consider Au + Au reactions at SIS energies
( about 1-2 GeV/nucleon ), since this
system has been extensively studied by the experimentalists
at GSI/Darmstadt. Secondly,
at an incident energy below the
the kaon production threshold
in free NN collisions ( about 1.58 GeV/nucleon for $K^+$ mesons
and 2.5 GeV/nucleon for $K^-$ mesons),
one can expect a low thermal velocity of the kaons. Consequently,
a collective motion, if it exists, will show up clearly.
In the present study we are especially interested in kaons emitted
at midrapidity.
These kaons are most likely produced in
compressed nuclear matter stopped in the overlap region of colliding
nuclei ( the so-called fireball ).
The reaction dynamics is thereby described within the transport
approach of Quantum Molecular Dynamics (QMD). 
We find in this work that
both the midrapidity
$K^+$ and $K^-$ mesons exhibit a novel collective motion in radial
direction, which we call kaon radial flow. We demonstrate that
the kaon radial flow can provide information on in-medium modifications
of kaon properties.
\vskip 0.5 true cm
This paper is organized as follows. In Section 2, we briefly sketch the
QMD approach and describe the treatment of the kaon dynamics.
In Section 3 we elaborate
the signature of the
$K^+$ radial flow and demonstrate its collective nature.
In Section 4, we study
the $K^-$ radial flow, which is found to be quite different from the
$K^+$ flow. The role of rescattering and reabsorption of the $K^-$ mesons
is also investigated. In Section 5 we summarize the main results
and conclusions of this work.
\vskip 0.5 true cm
\section{Description of kaons in QMD}
\subsection{QMD model}
The QMD model\cite{aichelin91}
has been extensively used for the investigation
of heavy ion reactions.
In addition to nucleons, we
include also $\Delta$ and $N^*$ resonances, which turned out to play
a non trivial role
at an incident energy of 1-2 GeV/nucleon\cite{hong97,hjort97}. Pions are produced
by the decay of these resonances. Their Coulomb interaction with the baryons
is also included. This is important for a
correct description of the pion dynamics, since the pion phase space is highly
isospin dependence\cite{bass94,ben79,uma97}.
 
\vskip 0.5 true cm
\subsection{Sources of $K^+$ and $K^-$ mesons}
In heavy ion reactions, $K^+$ and $K^-$ mesons are created from elementary
hadron-hadron collisions. Since only very energetic hadron-hadron
collisions are able to produce a kaon (or antikaon), one usually assumes that
no coherence exists between the different collisions producing kaons.
At SIS energies the major processes responsible for kaon
and antikaon production are baryon-baryon collisions
\beq
B + B \rightarrow B + Y + K^+
\eeq
\beq
B + B \rightarrow B + B + K^+(K^0) + K^-
\eeq
and pion-baryon collisions
\beq
\pi + B \rightarrow Y + K^+
\eeq
\beq
\pi + B \rightarrow B + K^+(K^0) + K^-.
\eeq
Here B stands for a nucleon or nucleonic resonance
while Y denotes a hyperon ( $\Sigma$ or $\Lambda$ ).
At SIS energies the relevant resonances are mainly  $\Delta$ and $N^*$
resonances. If one goes to higher
incident energies, e.g. to AGS energies ( about 10 GeV/nucleon ),
one must take into account contributions of heavier mesons and resonances
\cite{sorge91}.
For the baryon-baryon channels leading to $K^+$ or $K^-$ mesons, we have
adopted theoretical cross sections based on the one-boson-exchange model.
These cross sections agree well with experiments\cite{cassing97}.
The $K^+$ production cross section from the pion-induced channel
has been evaluated within a resonance model\cite{tsu94}, and a good agreement
has been found between theory and data.
The $K^-$ production cross section from pion-baryon
collisions is a parametrization of the experimental data\cite{cassing97}.
In a previous study, it was shown that at SIS energies pion-baryon reactions
can produce
even more $K^+$ mesons than baryon-baryon reactions\cite{fuchs97}.
With an increasing number of participating nucleons, the $K^+$ multiplicity originating
from the $\pi$-induced channel increases much faster than that from the
baryon-baryon channel. Hence this channel is essential in order to
reproduce the mass
dependence of the $K^+$ production observed in experiments\cite{miskowiec94}.
Furthermore, the pion-baryon reaction is
important if one wants to understand the experimental observation of
a non-isotropic $K^+$ angular distribution in symmetric heavy ion reactions
at subthreshold beam energies\cite{elmer96}. This is due to the fact that
the pion-baryon reaction of Eq.(3)
has a substantial P-wave component, which results in a large anisotropy.
The P-wave contribution turned out to be essential for the understanding
of the observed $K^+$ angular distribution\cite{wang972}.
\vskip 0.5 true cm
\subsection{Kaon reabsorption and rescattering}
Once produced a $K^+$ meson can be hardly absorbed inside the
medium because of strangeness conservation. However, a $K^-$ meson
can be easily destroyed through the reaction $K^-$ + N $\rightarrow$ $\pi$ + Y.
The inverse process also contributes partially to the $K^-$ production.
We used in our calculation a parametrization
of the experimental cross section for these reactions\cite{cassing97}.
The $K^-$ absorption is rather strong.
$\sigma_{K^-N \rightarrow \pi Y}$ is about 50 mb at a laboratory
$K^-$ momentum of 0.2 GeV/c and increases rapidly with decreasing momenta.
 
\vskip 0.5 true cm
Both $K^+$ and $K^-$ mesons may scatter elastically with
nucleons as propagating through the nuclear medium.
The $K^+$-nucleon interaction is relatively weak
( $\sigma_{KN}$$\sim$ 10 mb )\cite{dover82}. However,
in the initial stage of the nucleus-nucleus collisions at SIS energies,
one can expect
nuclear matter densites up to three times saturation density\cite{uma97}, which
considerably shortens the mean free path of the kaons ($\lambda$=
($\sigma$$\cdot$$\rho$)$^{-1}$). This may
lead to observable effects on the kaon spectra\cite{wang972,zwermann84}.
Compared to $K^+$ mesons, $K^-$ mesons have a much larger
elastic cross section,
which is about 50 mb at a laboratory $K^-$
momentum of 0.2 GeV/c and becomes even larger with decreasing momenta.
For the elastic $K^+$ and $K^-$ cross sections we use the parametrization
given in Ref.\cite{cassing97}.
\vskip 0.5 true cm
\subsection{Kaon Coulomb interaction}
In Ref.\cite{wang97} it was shown that the Coulomb interaction
between the $K^+$ mesons and the nuclear medium has a visible effect
on the kaon dynamics in heavy systems like Au + Au.
Thus the Coulomb interaction of
both the $K^+$ and $K^-$ mesons is taken into account.
The Coulomb field
acting on the kaons is generated by protons, charged resonances and
pions. We should notice that the multiplicity of negative charged pions
has been observed to exceed greatly the value for positive
charged pions in large systems such as Au + Au\cite{bass94,ben79,uma97}.
This leads to a non-vanishing contribution
of the pions to the charge density. Thus the total Coulomb interaction
is given by the sum over all charged particles, i.e.
\beq
U^{Cou}_{i_k} = \sum_{i_n}\frac{Z_{i_k}Z_{i_n}e^2}
{\mid{{\vec r}_{i_k}-{\vec r}_{i_n}}\mid}+\sum_{i_r}\frac{Z_{i_k}Z_{i_r}e^2}{\mid{{\vec r}_{i_k}-{\vec r}_{i_r}}\mid}
+\sum_{i_\pi}\frac{Z_{i_k}Z_{i_\pi}e^2}{\mid{{\vec r}_{i_k}-{\vec r}_{i_\pi}}\mid},
\eeq
where $i_n$, $i_r$ and $i_\pi$ denote nucleons, resonances
($\Delta$ and $N^*$) and pions, respectively.

\vskip 0.5 true cm
\subsection{Kaon mean field of the strong interaction}
The strong interaction of kaons with nuclear matter is expected to play also
an important role in modifying the propagation of $K^+$ and $K^-$ mesons.
Possible effects of the kaon-nuclear matter
strong interaction have attracted special interest\cite{li95,wang97,ritman95,barth97,cassing97}.
The well-established formalism for studying kaon properties in
nuclear matter is chiral perturbation theory. Based on a SU(3)$_L$$\times$SU(3)$_R$
chiral Lagrangian Kaplan and Nelson\cite{kaplan86} constructed an effective
meson-baryon Lagrangian, whose coefficients had been determined from
experimental measurements, however, still with an uncertainty of about 30$\%$.
Applying
this Lagrangian to the KN interaction on the mean field level, one can
obtain the following dispersion relation for a kaon or antikaon in nuclear matter\cite{li95}
\beq
\omega_{K^+} = \sqrt{\vec{p}^2 + m_K^2[ 1 - \frac{\Sigma_{KN}}{f_{\pi}^2m_K^2}\rho_S
+ (\frac{3\rho_B}{8f_\pi^2m_K})^2]} + \frac{3}{8}\frac{\rho_B}{f_{\pi}^2},
\eeq
\beq
\omega_{K^-} = \sqrt{\vec{p}^2 + m_K^2[ 1 - \frac{\Sigma_{KN}}{f_{\pi}^2m_K^2}\rho_S
+ (\frac{3\rho_B}{8f_\pi^2m_K})^2]} - \frac{3}{8}\frac{\rho_B}{f_{\pi}^2},
\eeq
where $m_K$ denotes the bare kaon mass, and $f_\pi$ $\approx$ 93 MeV
is the pion decay constant. $\rho_B$ and $\rho_S$ are
the baryon density and the scalar density, respectively.
The kaon-nucleon sigma term $\Sigma_{KN}$
depends on the nucleon strangeness content and the strange quark mass.
A value of $\Sigma_{KN}$ = 350 MeV has been used in the literature\cite{li95,cassing97}, however, recent lattice
QCD calculations imply a higher value of $\Sigma_{KN}$ = 450 $\pm$ 30 MeV\cite{brown96}.
While the $\Sigma_{KN}$ term provides
a scalar attraction for both $K^+$ and $K^-$ mesons, the term
proportional to $\rho_B$ gives rise
to a vector potential, which is repulsive for $K^+$ mesons and attractive
for $K^-$ mesons due to the G-parity transformation.
The expressions
Eq.(6) and Eq.(7) for the in-medium kaon potentials include only
the S-wave interaction. There is also a P-wave contribution of kaons
with the nuclear medium by hyperon-nucleon hole excitations.
This P-wave interaction changes signs from $K^+$ to $K^-$ mesons
and is in its absolute value about twice as large for $K^-$ than for
$K^+$ mesons. This P-wave interaction has till now been neglected
in all previous works on kaons in nuclei. We will neglect it
here also. But future investigations should study, if it is important.
The strong potential for a kaon is in this approach defined as\cite{shuryak92}
\beq
U_K^{Str} = \omega_K - \sqrt{m_K^2 + \vec{p}^2}.
\eeq
Considering the mean field of the strong interaction and the
Coulomb field, the total Hamiltonian for the kaons reads
\beq
H_k = \sum_{i_k}\left (\sqrt{m_K^2+{\vec P}_{i_k}^2}+U^{Cou}_{i_k}+
U^{Str}_{i_k}\right ),
\eeq
\vskip 0.5 true cm
There exist experimental constraints on the kaon mean field.
As indicated above, the $K^+$N interaction is relatively weak compared
to other hadron-nucleon interactions. Therefore, the impulse approximation
should be justified to determine the in-medium $K^+$ potential
at low densities from experimental free KN scattering data.
This yields a repulsive $K^+$ potential of about 30 MeV
at saturation
density ($\rho_0$ = 0.16 $fm^{-3}$), if one adopts an isospin
averaged free KN scattering length $\bar{a}_{KN}$ =  -0.255 fm\cite{barnes94}.
On the other hand, experiments of kaonic atoms can provide direct
information on the $K^-$ mean field around $\rho_0$. An attractive
$K^-$-nucleus potential
of $U_{K^-}$ = -200 $\pm$ 20 MeV had been found
in $^{56}$Ni\cite{batty97}. With $\Sigma_{KN}$ = 350 MeV, one obtains from
Eq.6-8 a $K^+$ potential of about +7 MeV and a $K^-$
one of -100 MeV at $\rho_0$. Both values are
quite different from their respective empirical findings. High order
corrections and P-wave contributions to the mean field approximation may be responsible for
this discrepancy. In Ref.\cite{brown96,brown972} it was argued
that the so-called range term, which is of the same order in chiral
perturbation theory as the $\Sigma_{KN}$ term, has also contributions.
In addition, the presence of the nuclear medium
can reduce the pion decay constant
$f_{\pi}$ by decreasing the quark condensates. Taking the range term
and the in-medium reduction of $f_{\pi}$ into account,
one obtains at $\rho_0$ about +28 MeV for the $K^+$ and -175 MeV
for the $K^-$ potential with a value of $\Sigma_{KN}$ = 450 MeV.
Thus the strength of the potentials is now roughly consistent
with the respective empirical values.
 
\vskip 0.5 true cm
The $K^-$-nuclear matter interaction is complicated by the
$\Lambda$(1405) resonance in the $K^-$-nucleon channel,
which is commonly considered to be
an unstable $\bar{K}$N I = 0 bound state about 27 MeV below
the $K^-$p threshold. This state also appears as a resonance in the
$\pi$$\Sigma$ channel.
The $\Lambda$(1405) is responsible for the repulsive
isospin-averaged
$\bar{K}$N scattering length in free space\cite{batty97}.
Since the $\bar{K}$N interaction is strong, the impulse
approximation should not be good. A coupled channel description of
the $\bar{K}$N and $\pi$-hyperon channels is necessary.
Such studies including the $\Lambda$(1405)
have been performed by Weise et al.\cite{kaiser95,waas96}.
They show that the $\Lambda$(1405) state dissolves
in a nuclear environment due to Pauli blocking.
Therefore, although the $\Lambda$(1405) dominates the
threshold $\bar{K}$N dynamics at very low densities ($\rho$ $<$ 0.3$\rho_0$),
it is of minor relevance for the $K^-$-nuclear matter interaction
at densities $\rho$ $>$ $\rho_0$. These calculations\cite{kaiser95,waas96}
have been performed with free kaons. A selfconsistent inclusion of
the kaon-nucleus potential (in medium effects) may change these conclusions and
should in the future be considered.

\vskip 0.5 true cm
For the $K^+$ mesons, we use the potential suggested in Ref.\cite{brown96,brown972}, since
it agrees with the impulse approximation.
For $K^-$ mesons, a potential
of -200 MeV at $\rho_0$ seems to be too deep to describe
the experimental data of $K^-$ production from heavy ion
reactions\cite{cassing97,li972}.
Thus one usually uses a $K^-$ potential of about -100 MeV at normal
nuclear matter density\cite{cassing97,li972}.
We use a $K^-$
potential derived from Eq.7-8 with $\Sigma_{KN}$ = 350 MeV.
The additional corrections from P-wave contributions, the range term and the in-medium
modification of the pion decay constant are thereby neglected.
Such a $K^-$ potential is more or less
the same as used by other authors\cite{cassing97,li972}, and
agrees reasonably well with coupled
channel studies\cite{kaiser95,waas96}. We will show later that the conclusions
of the present work do not depend on the particular choice of the $K^-$ potential.
However, there seem to exist some discrepancies between the $K^-$ potential extracted from the kaonic
atoms and the one required for understanding $K^-$ production from
heavy ion reactions\cite{cassing97,li972}. Future studies should shed light on
this question.
In Fig. 1 the corresponding potentials for $K^+$ and
$K^-$ mesons used in the present work are shown for zero momentum.
We note that the scalar density $\rho_S$ is model dependent.
We followed a method used in the RBUU ( Relativistic 
Boltzmann-Uehling-Uhlenbeck ) model to calculate $\rho_S$
as a function of $\rho_B$\cite{weber92}.
\vskip 0.5 true cm
The kaon production thresholds might also be changed
in the nuclear environment by
the strong mean field $U^{Str}$. ( We note that the Coulomb field
$U^{Cou}$ has no net effect on the threshold due to charge conservation. )
The KaoS Collaboration\cite{barth97} has observed
an enhanced $K^-$ production
in nucleus-nucleus collisions as compared to free nucleon-nucleon
collisions, which could
be attributed to an attractive $K^-$ mean field.
Thus, we have taken into account the reduction of the $K^-$
production threshold in a way similar as in Ref.\cite{cassing97}.
The $K^+$ mean field is much weaker
than the $K^-$ one.
About medium modifications of the $K^+$ production threshold
exist different opinions
in the literature\cite{cassing97,brown96,fang94}. 
If such an effect exists, it should mainly affect the multiplicity
of the kaons, rather than their collective motion. Therefore,
we have not included any medium modification of the $K^+$ production
threshold in the present work, since the
$K^+$ mean field is much weaker than that for the $K^-$ mesons.
One can expect that the threshold is also much less affected.
\vskip 0.5 true cm
The QMD model described above has been shown to be able
to reproduce very well the
experimental measurements of the multiplicities and transverse flow of
nucleons, pions and kaons\cite{wang97,fuchs97,wang972,uma97}. In this work, we will apply the same model
to other kaon collective flows (e.g. radial kaon flow).

\vskip 0.5 true cm
\section{$K^+$ radial flow}
\subsection{Transverse mass spectra}
One way to obtain information on the collective motion in heavy ion reactions
is to investigate particle multiplicities as a function of the transverse
mass ( $m_t$ = $\sqrt{m^2_0 + P^2_T}$ ).
In Fig. 2 we show the transverse mass spectrum of $K^+$ mesons emitted at midrapidity 
( -0.4$<$$y_{c.m.}$/$y_{proj}$$<$0.4 ) in
the reaction Au + Au at an incident energy of 1 GeV/nucleon and
an impact parameter of 3 and 5 fm, respectively. A striking finding
of the present work is that the kaon spectrum
exhibits as a function of the transverse mass a "shoulder-arm" shape,
which deviates from a pure thermal picture.
This becomes evident by comparing the kaon spectrum with a Boltzmann fit
to the high energy part of the spectrum ( also shown in Fig. 2 ). 
In a pure thermal picture
the kaons are produced from an equilibrated thermal source.
The transverse mass spectra should thus satisfy a Boltzmann
distribution, which is
more or less a straight line if plotted logarithmically. 
A similar deviation from the thermal behavior had already been found
for nucleons and composite particles ( deuterons, heliums etc. ),
which was explained by introducing a collective expansion
of the fireball, i.e. the nuclear matter radial flow\cite{siemens79,matt95,danielewicz95}.
This expanding
source picture has also been successfully applied to pion and kaon spectra
at AGS energies ( about 10 GeV/nucleon )\cite{lee90}. This seems to be reasonable since at such high beam energies
hadron-hadron collisions occurring in the late expansion phase of the reaction
are still energetic enough for kaon production. The kaons produced in the
expanding nuclear matter should reflect a collective component
transfered from the nucleons to kaons.
However, it is not clear if this mechanism still works
at subthreshold energies.

\vskip 0.5 true cm
In Fig. 3 calculations are presented where the $K^+$ potential
$U^{Str}$ and
the Coulomb interaction $U^{Cou}$ have been dropped.
Now one sees only effects
of the hadron-hadron collisions, which produce the kaons.
( The $K^+$-nucleon elastic scattering is also included in the simplified
calculations, however, its influence is fairly weak. )
The resultant kaon transverse mass spectrum with
the same rapidity cut as in Fig. 2 is in exact agreement with a Boltzmann distribution.
Thus, we can deduce that the "shoulder-arm" structure observed in Fig. 2
is caused by the kaon mean field and the Coulomb
interaction, rather than by a collective expansion of the kaon sources.
As pointed out in Ref.\cite{wang97},
the mean field and not the Coulomb interaction
plays the dominant role in modifying the kaon motion.
The fact that the potential and not an expanding $K^+$ source is responsible for the
"shoulder-arm" structure can be connected with the subthreshold beam energy.
Only the hadron-hadron collisions
during the early compression phase of the reaction
have enough energy to produce a kaon. As the fireball
starts to expand coherently in a later stage,
the kaon production
ceases to stop since the energetic collisions have been exhausted. 
Consequently, at a subthreshold beam energy no "shoulder-arm" spectrum
for the midrapidity kaons can arise from the velocities of the sources.
On the other hand, the $K^+$ mean field is repulsive
with a value of about 28 MeV at saturation density ( $\rho_0$ ) and increases with
the density.
The kaons will thus inevitably experience an acceleration due to the
repulsive potential
as they propagate outwards from the overlap region.
This may give rise to a collective motion of
the kaons in the radial direction, which can lead to a deviation from the
thermal picture. Therefore, the
observed "shoulder-arm" structure in the kaon spectrum appears to be
a signature of a genuine collective
motion of the kaons, rather than only a result
of the nuclear matter radial flow.

\vskip 0.5 true cm
\subsection{Radial flow}
In order to quantitatively extract the collective component from the kaon spectrum,
we fit the QMD results by incorporating a common radial
velocity of the $K^+$ mesons to the standard Boltzmann distribution. The
corresponding distribution reads
\beq
\frac{{d^3}N}{d{\phi}dy{m_t}d{m_t}} \sim e^{-(\frac{{\gamma}E}{T}-{\alpha})}
\{ {\gamma}^{2}E - {\gamma}{\alpha}T(\frac{E^2}{p^2}+1) + ({\alpha}T)^{2}\frac{
E^2}{p^2} \}\frac{\sqrt{({\gamma}E-{\alpha}T)^{2}-m^{2}}}{p}
\label{reswidth}
\eeq
where $E$ = $m_t$$coshy$, $p$ = $\sqrt{{p_t}^2+{m_t}^2sinh^2y}$,
$\alpha$ = $\gamma$$\beta$$p/T$, $\gamma$ = $(1-\beta^2)$$^{-1/2}$.
T is the temperature of the thermal source,
while the parameter $\beta$ = $\upsilon$/c is a common
radial velocity of the kaons.
Fig. 2 shows that Eq.(10) gives a good fit to the QMD results.
The fit yields for b = 5 fm a temperature of T = 60 MeV and a common velocity of
$\beta$ = 0.1. A slightly higher temperature ( T = 62 MeV ) and larger
common velocity ( $\beta$ = 0.11 ) are obtained for more
central collisions ( b = 3 fm ).
Are the T and $\beta$ derived from the fit a good measure of
the thermal and collective motion of the kaons? As we
mentioned above, the thermal feature of the kaon spectra described by T
reflects the temperature of
the kaon sources, while
the collective component - given by $\beta$ - is due to the acceleration
 of the kaons
by the mean field. Thus the extracted temperature describes
the thermal motion without the influence of the
kaon potential ( see Fig. 3 ). On the other hand,
the collective motion $\beta$ can be quantitatively described by the enhancement
of the kaon momentum by the repulsive $K^+$ potential.
The simulations shown in Fig. 3 give temperatures of
T = 64 MeV for b = 5 fm and T =
66 MeV for b = 3 fm, respectively. They are quite comparable to the
temperatures T extracted from the fit to the full calculations (Fig. 2).
The average transverse mass of the kaons is
increasing from $<$$m_t$$>$ = 566 MeV to 594 MeV for b = 5 fm as one changes
from the simulations without to those including the kaon potential.
For b = 3 fm it increases from $<$$m_t$$>$ = 571 MeV to 600 MeV. ( We note that
the Coulomb interaction contributes less than
30$\%$ to the enhancements. ) 
This corresponds
to a common radial velocity boost of the kaons by $\beta$ = 0.102 and
$\beta$ = 0.113 for b = 5 fm and 3 fm, respectively. The two values are
very close to the $\beta$ parameters obtained by the fit.
This indicates that it is possible to separate the thermal
and the collective components of the kaon spectrum by the present
analysis.

\vskip 0.5 true cm
At an incident energy of 1 GeV/nucleon,
the energy available for kaon production is quite limited. Consequently,
the kaons have also a limited thermal motion.
The average radial velocity of the
kaons due to thermal motion is about 0.488 ( in unit of the light velocity c )
for Au + Au, while the flow velocity is about $\beta$ = 0.1 as mentioned above.
The collective motion is more than 20$\%$ of the thermal motion,
so that it is not obscured by the latter, and
can be observed as a clear signal.

\vskip 0.5 true cm
The radial collective motion of the $K^+$ mesons
may provide a new opportunity for investigation of the kaon
mean field in nuclear matter.
As stronger the kaon potential, as more significant
is the radial kaon flow. Thus the kaon radial flow can be
used as a meter to determine the kaon in-medium potential.
We note that
an analysis of the transverse momentum of the kaons as 
a function of the rapidity (so-called kaon transverse flow) might
give also information of the kaon potential.
Without the kaon mean field, the transverse kaon flow
follows the flow patterns of the baryons, and exhibits a S-shape.
This has been found in both RBUU calculations\cite{li95} and
previous studies using the QMD model\cite{wang97}.
The S shape disappears and a zero flow can be observed if one
takes into account the kaon potential as found in Ref.\cite{li95,wang97,ritman95}.

\vskip 0.5 true cm
\section{$K^-$ "virtual" radial flow}
In contrast to the $K^+$ mesons studied in the preceding section,
many other hadrons such as $K^-$, $\Lambda$, $\Sigma$ etc.
are supposed to feel an attractive mean field in nuclear matter.
In addition they may be strongly absorbed or scattered in the medium.
If these hadrons can also
develop a radial collective motion in heavy ion reactions is
not clear. Therefore, it is of high
interest to investigate the possible formation of radial flow of these hadrons.
As a first example, we consider in the present work $K^-$ mesons which are
subjected to all the final-state interactions mentioned above.
\vskip 0.5 true cm
In Fig. 4 the $K^-$ transverse mass spectrum
at midrapidity is shown for the Au + Au reaction
at a single impact parameter b = 5 fm and for the Ni + Ni reaction by
integrating over impact parameters 0$<$ b $<$4 fm.
The considered beam energies ( 1.8 GeV/nucleon for the Au + Au reaction
and 1.93 GeV/nucleon for the Ni + Ni reaction )
are below the free NN $K^-$ threshold.
We analysed the $K^-$ mesons from a rapidity interval
( -0.3 $<$ $y_{c.m.}$/$y_{proj}$ $<$ 0.3 ) normalized to projectile
rapidity $y_{proj}$, which looks
narrower than the one for $K^+$ mesons
( -0.4 $<$ $y_{c.m.}$/$y_{proj}$ $<$ 0.4 ). In fact, the size
of the intervals in units of rapidity is very close to each other,
since
we consider for the $K^-$ production higher beam energies than
for the $K^+$ production due to the different thresholds ( 2.5
GeV compared to 1.58 GeV ).
The rapidity of the $K^+$ or $K^-$ mesons
is explicitly included in the equations fitting the kaon
spectra ( see Eq.10 for the $K^+$ case and Eq.11 to appear later
in this section for the $K^-$ case ). Therefore, small differences between
the rapidity intervals for the $K^+$ and $K^-$ mesons should not influence
the conclusions of this study.
Also shown in Fig.4 is a Boltzmann fit to the high energy part of the spectrum
( $m_T$ - $m_{K^-}$ $>$ 0.1 GeV ).
One can clearly observe a "concave" structure in the
$K^-$ spectrum again deviating from a thermal Boltzmann distribution.
This structure is, however, completely different from the "shoulder-arm" shape found for $K^+$ mesons.

\vskip 0.5 true cm
The "concave" structure is a consequence of the attractive $K^-$ mean field.
Without the mean field this structure vanishes.
( The present work also includes the Coulomb
interaction of the $K^-$ mesons. However, the Coulomb effects are found
to be negligible compared to those originating from
the strong potential. )
Fig. 5 demonstrates that
the $K^-$ spectra obtained without the
mean field and the Coulomb interaction
( but including rescattering and reabsorption ) are in good
agreement with a Boltzmann distribution. At a subthreshold beam energy
hadron-hadron collisions, i.e. sources
for the $K^-$ mesons,
can not induce this "concave" structure
since the $K^-$'s are mainly produced before
the fireball starts to expand.
We already discussed this point in Section III.
It is important to realize that also reabsorption and rescattering
do not lead to the observed "concave" structure.
The transverse mass spectrum of the midrapidity $K^-$ mesons is determined
primarily by the interaction of the $K^-$ mesons with the fireball.
If one neglects all the final-state
interactions, the primordial $K^-$ spectra should satisfy
according to Fermi's argument\cite{fermi50} a pure Boltzmann distribution.
On the other side,
the fireball baryons show predominantly a single-temperature structure
before the formation of the nuclear matter radial flow.
We know that both rescattering and reabsorption are in nature stochastic
processes which tend to thermalize a narrow distribution.
One can hardly expect that a stochastic interaction between two
thermal distributions can change the thermal distributions to a two-temperature
structure. Thus one observes in Fig. 5 a Boltzmann distribution of the $K^-$ mesons even though rescattering and reabsorption have been included. However,
we should note that
rescattering and reabsorption
of hadrons by spectators can induce non-thermal
features since the spectators are collectively
deflected sidewards in the reaction plan ( nuclear matter bounce-off ).
Absorption of pions by the spectator baryons is responsible
for a pion transverse flow which is anticorrelated to the nucleon
flow\cite{bli91,bass932,kint97}.
Rescattering and reabsorption
by the spectators can also give rise to an azimuthal anisotropy in
the hadron emission (Squeeze-out)\cite{brill93,venema93,bass94}.
\vskip 0.5 true cm
In Section III we have shown that the $K^+$ radial flow can be described very well by
boosting a Boltzmann distribution with a common $K^+$ radial velocity
directed away from the fireball as a result of the repulsive
$K^+$ mean field. Here we try to understand the $K^-$ spectra with
a similar boost, however, now with a common $K^-$ radial velocity tending
towards
the center of the fireball since the $K^-$ mesons feel an attractive
mean field. In analogy to Eq.(10) the $K^-$ distribution reads

\beq
\frac{{d^3}N}{d{\phi}dy{m_t}d{m_t}} \sim e^{-(\frac{{\gamma}E}{T}+{\alpha})}
\{ {\gamma}^{2}E + {\gamma}{\alpha}T(\frac{E^2}{p^2}+1) + ({\alpha}T)^{2}\frac{
E^2}{p^2} \}\frac{\sqrt{({\gamma}E+{\alpha}T)^{2}-m^{2}}}{p}.
\label{reswidth}
\eeq
T is the primordial $K^-$ temperature, $\beta$ is the common $K^-$ velocity.
Fig. 4 shows that Eq.(11) provides a good fit to the $K^-$ spectrum obtained from the
full calculation.
The adjusted temperature parameters ( T = 83 MeV and 76 MeV
for Au + Au and Ni + Ni, respectively ) are again close to these values deduced from
a Boltzmann fit to the QMD results without the mean field ( T = 87 MeV and
82 MeV for the two reactions, respectively ). This indicates that the fitting
parameter T in Eq.(11) is a good measure for the thermal motion of the
$K^-$ mesons. The average transverse mass of the $K^-$ mesons is found
to be reduced by the mean field from $<m_T>$ = 596 MeV
to 575 MeV for Au + Au, and from 585 MeV to 569 MeV for Ni + Ni, which
correspond to a radial decelerating boost of $\beta$ = 0.083 and 0.067,
respectively.
Both $\beta$ values agree with that from the fit ( $\beta$ =
0.085 for Au + Au and 0.066 for Ni + Ni ). Therefore, the
non-thermal feature of the $K^-$ mesons can be described well by
a radial velocity $\beta$ extracted from the fit.

\vskip 0.5 true cm
The deviation of the $K^-$ transverse mass spectra from a
pure thermal distribution can therefore be understood in terms of a common
$K^-$ radial collective motion, which
arises from the deceleration by the attractive mean field.
We call this type of collective motion "virtual" flow
in order to distinguish it from the $K^+$ radial flow driven by a
repulsive mean field.
Such a virtual flow is characterized by a "concave" structure in the transverse
mass spectrum, while a radial flow shows up
in a "shoulder-arm" shape of the spectrum.

\vskip 0.5 true cm
Rescattering and reabsorption alone can not induce
the "concave" feature in the $K^-$ spectra.
However, a stochastic process may interfere with a collective motion
as mentioned above.
Reabsorption plays for $K^-$ mesons a larger role than for $K^+$
mesons. $K^-$ rescattering on the fireball
baryons may increase the energy of $K^-$ mesons. This will
cancel partially the virtual flow caused by the mean field.
Therefore, we observe in the present work a relative small reduction
of the $K^-$ transverse mass ( $\Delta$$<m_T>$ = - 21 MeV and - 16 MeV
for Au + Au and Ni + Ni, respectively. ) although we have adopted a
mean field for the $K^-$-nuclear matter interaction of about -100 MeV at normal nuclear matter density.

\vskip 0.5 true cm
A "concave" structure had been observed
experimentally in the transverse mass spectra for $K^+$ and $K^-$ mesons
in Si + Pb at 14.6 (GeV/c)/nucleon\cite{stachel94}
and Au + Au at 10.8 (GeV/c)/nucleon\cite{barrette95}
as well as for pions in many reactions and at various beam energies\cite{brockmann84}. In all the
cases the beam energies are much higher than the corresponding production
thresholds for free NN collisions. Considerable numbers of
kaons or pions
are produced even after the fireball starts to expand.
Thus the fireball expansion
can independently cause non-trivial collective features in the particle
spectra\cite{danielewicz95,lee90}. Resonance decay producing pions has also a significant
contribution, which had been found
to be able to induce a "concave" structure in pion transverse mass spectra\cite{hong97,hjort97,sorge89}.
It is worth noting
that some explanations for these observations have also used
an attractive pion or kaon
in-medium mean field\cite{xiong93}. However, it seems that
there exist some differences between these
observations and our present work where the subthreshold beam energy plays
an essential role.

\vskip 0.5 true cm
Finally we want to add some remarks concerning the kaon potentials
used in this work. For the description of the $K^+$ radial flow
a potential has been used which agrees with the free KN scattering data.
Based on the same potential, we were able to reproduce the experimental
$K^+$ transverse flow data in a previous work \cite{wang97}.
It is also unlikely for resonance decay to modify the "shoulder-arm" shape
of the $K^+$ spectra, since, at the incident energies studied here,
very few Phi mesons and antibaryons can be created. The $K^-$
potential applied is less attractive than that extracted from
kaonic atoms. Since the deceleration of the
$K^-$ mesons by the attractive mean field is responsible for the
"concave structure" of the $K^-$ spectra, a more attractive potential
should even favour a more pronounced "concave" shape. 
The $\Lambda$(1405)
might also play a role in leading to a "concave" structure of the $K^-$ spectra
just analogous to the role of the $\Delta$ resonances
for pions. However, the formation of the $\Lambda$(1405) is reduced
at high densities. Since the $K^-$ radial
flow arises from interactions with the dense fireball,
one can expect that the $\Lambda$(1405) is only of minor importance.
We also note that our treatment of the kaon dynamics is obviously
non-covariant, since e.g. the space component of the vector
part of kaon-nucleus potential is neglected as usual\cite{fuchs98}.
However, this treatment is sufficient
for the present
studies, since the relative motion between the fireball and
the midrapidity kaons is quite small
at the subthreshold beam energies. The present conclusions are not
significantly affected by a full covariant treatment of the kaon
dynamics.
\vskip 0.5 true cm
\section{Summary}
In this work, we have studied possible collective motion of $K^+$ and
$K^-$ mesons emitted at midrapidity from heavy ion reactions
at SIS energies (1-2 GeV/nucleon) by comparing with a thermal
picture. The main findings are:
\vskip 0.5 true cm
(1). A collective flow of $K^+$ mesons in the radial direction
should exist in central nucleus-nucleus collisions, which can
be identified by a characteristic "shoulder-arm" shape
of the transverse mass spectrum of the midrapidity kaons. The $K^+$
radial flow is caused by the repulsive $K^+$ mean field.
\vskip 0.5 true cm
(2). $K^-$ mesons also exhibit a collective
motion in the radial direction in heavy ion reactions at subthreshold beam
energies even though they suffer strong reabsorption and rescattering
in the nuclear medium. The collective motion leads to a characteristic
"concave" structure in the transverse mass spectrum of $K^-$ mesons emitted at
midrapidity. This distinguishes the $K^-$ collective motion
from that of $K^+$ mesons.
The $K^-$ collective motion is induced by an
attractive mean field.
\vskip 0.5 true cm
(3). Since the radial flow of both $K^+$ and $K^-$ mesons arises from
their respective in-medium mean field, the radial flow may
act as a novel tool for investigation of
possible modifications of kaon properties in dense nuclear matter.
\vskip 0.5 true cm
\vskip 0.5 true cm
\vskip 0.5 true cm
\vskip 0.5 true cm
\vskip 0.5 true cm
\vskip 0.5 true cm
\newpage

\newpage
Figure Captions
\vskip 0.5 true cm

Fig. 1. The in-medium potentials of $K^+$ and $K^-$ mesons in nuclear matter
used in this work.

\vskip 0.5 true cm

Fig. 2. Transverse mass spectra of the $K^+$ mesons emitted at
midrapidity ( -0.4$<$$y_{c.m.}$/$y_{proj}$$<$0.4 ).
The histograms are results of the QMD simulations for Au+Au reactions
at 1 GeV/nucleon and at two different impact parameters b = 3, 5 fm. The
calculations are performed including the full in-medium kaon dynamics.
The lines are fits to the transport calculations. In the upper panel
the result of a pure thermal Boltzmann fit to the high energy part of the
spectra ( $m_t$-$m_0$ $>$ 0.15 GeV ) is shown. In the lower panel
the fit according to Eq.(10) is presented where one assumes a $K^+$ radial
flow in addition to the thermal motion.

\vskip 0.5 true cm
Fig. 3. Transverse mass spectra of the $K^+$ mesons emitted at
midrapidity ( -0.4$<$$y_{c.m.}$/$y_{proj}$$<$0.4 ) for the same
reactions as in Fig. 2. The histograms present results of the
transport calculations without the $K^+$ mean field and Coulomb
potential. Now a pure thermal fit (lines )
is sufficient to reproduce the spectra.

\vskip 0.5 true cm
Fig. 4. Transverse mass spectra of the $K^-$ mesons emitted at
midrapidity ( -0.3 $<$ $Y_{c.m.}$/$Y_{proj}$ $<$ 0.3 ) in a semicentral
( b = 5 fm ) Au + Au reaction at 1.8 GeV/nucleon and in a Ni + Ni reaction at
1.93 GeV/nucleon. The Ni + Ni reaction is averaged over the impact parameter region
0 $<$ b $ <$ 4 fm.
The histograms are the results of the transport calculations
including the full $K^-$ in-medium dynamics and the lines
are the corresponding fits to the spectra.
In the upper panel
the result of a pure thermal Boltzmann fit to the high energy part of the
spectra ( $m_t$-$m_0$ $>$ 0.1 GeV ) is shown.
In the lower panel
the fit according to Eq.(11) is presented where one assumes a $K^-$
"virtual" radial flow in addition to the thermal motion.

\vskip 0.5 true cm
Fig. 5. Transverse mass spectra of the $K^-$ mesons emitted at
midrapidity ( -0.3 $<$ $Y_{c.m.}$/$Y_{proj}$ $<$ 0.3 ) for the same
reactions as in Fig. 4. 
The histograms present results of the transport calculations without
the $K^-$ mean field and Coulomb potential.
Now a pure thermal fit (lines )
is sufficient to reproduce the spectra.
\newpage
\begin{figure}
\leavevmode
\epsfxsize = 12cm
\epsffile[60 85 430 750]{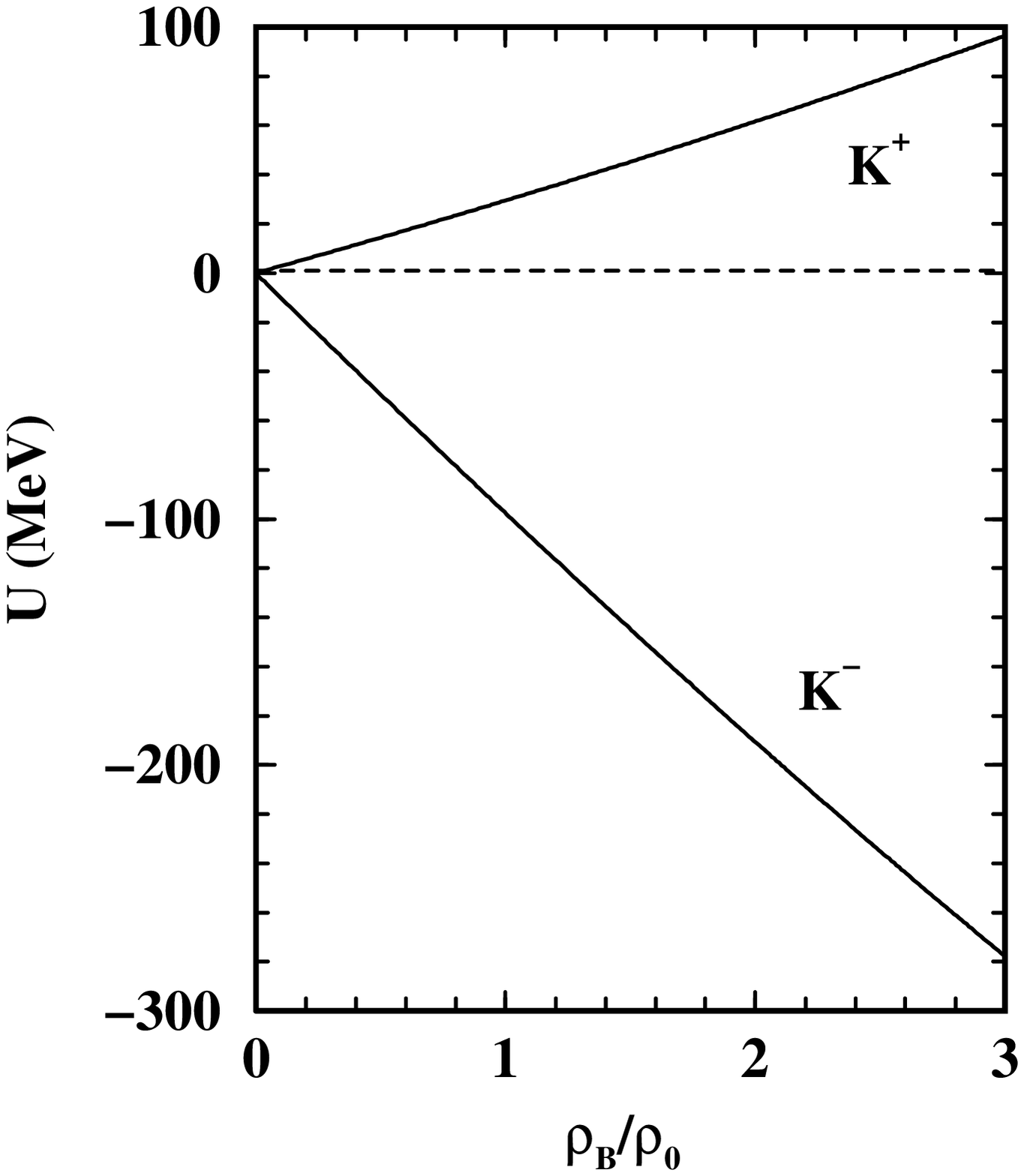}
\caption{
}
\label{fig2}
\end{figure}
\newpage
\begin{figure}
\leavevmode
\epsfxsize = 12cm
\epsffile[60 85 430 750]{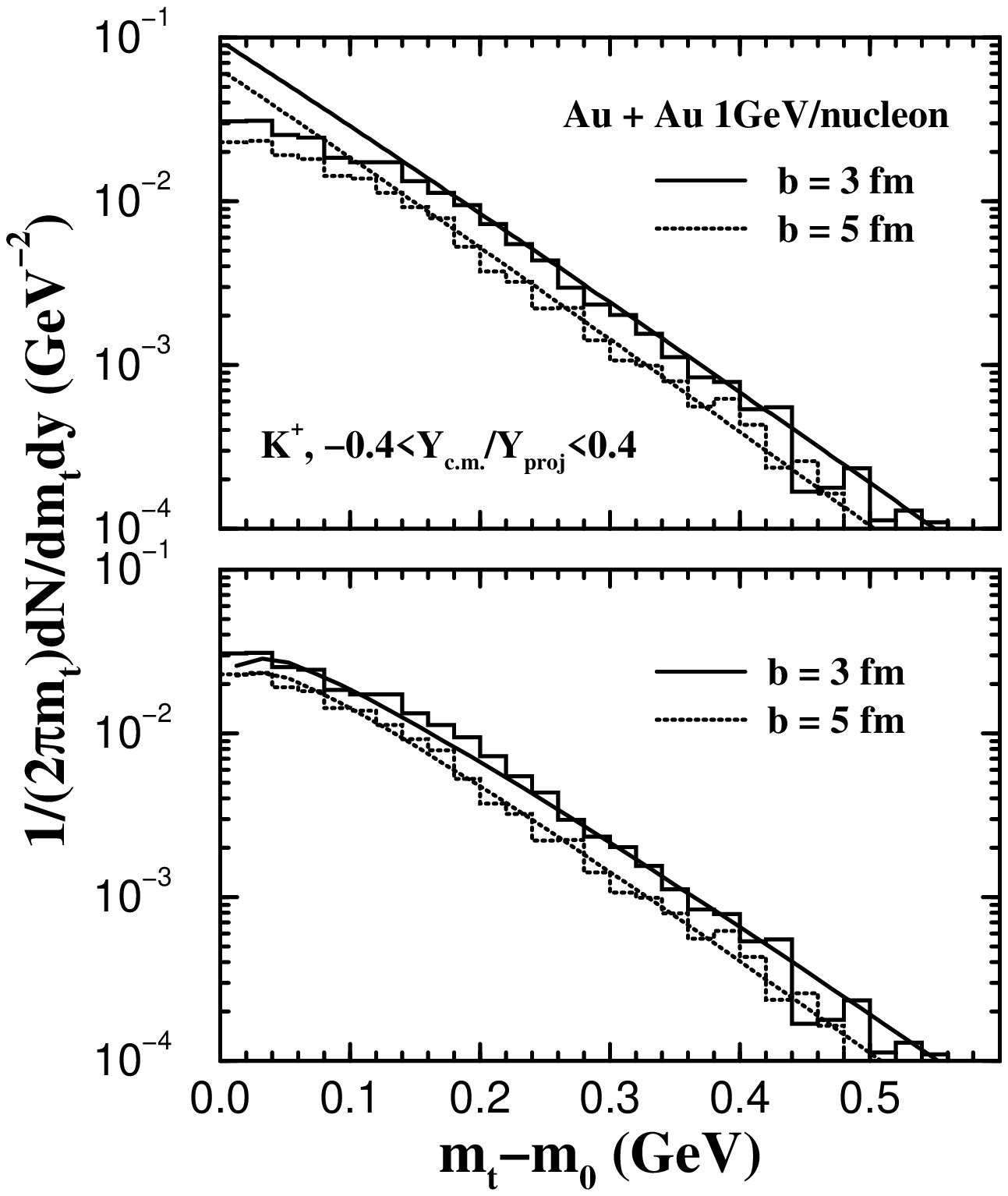}
\caption{
}
\label{fig2}
\end{figure}
\newpage
\begin{figure}
\leavevmode
\epsfxsize = 12cm
\epsffile[60 85 430 750]{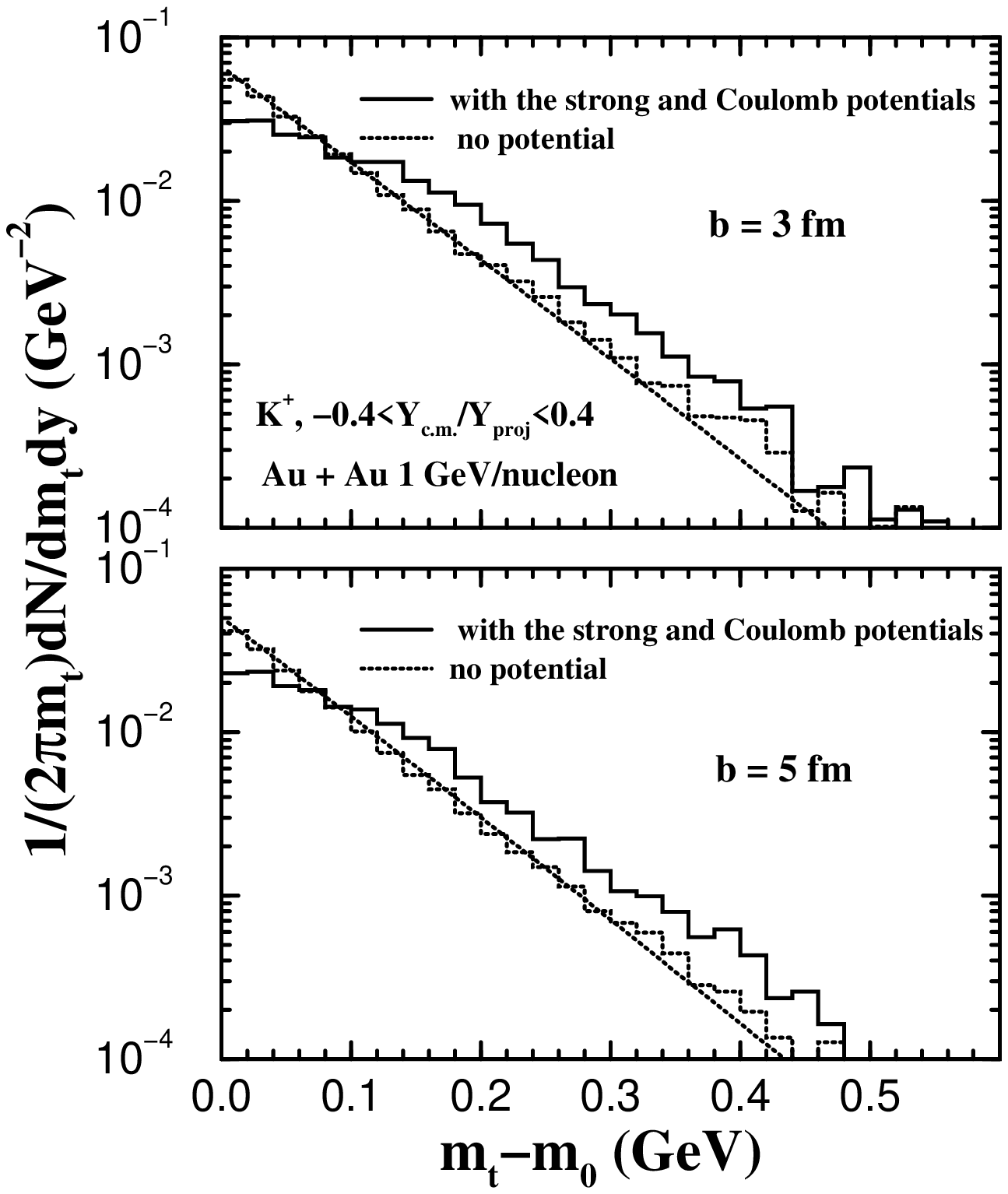}
\caption{
}
\label{fig3}
\end{figure}
\newpage
\begin{figure}
\leavevmode
\epsfxsize = 12cm
\epsffile[60 85 430 750]{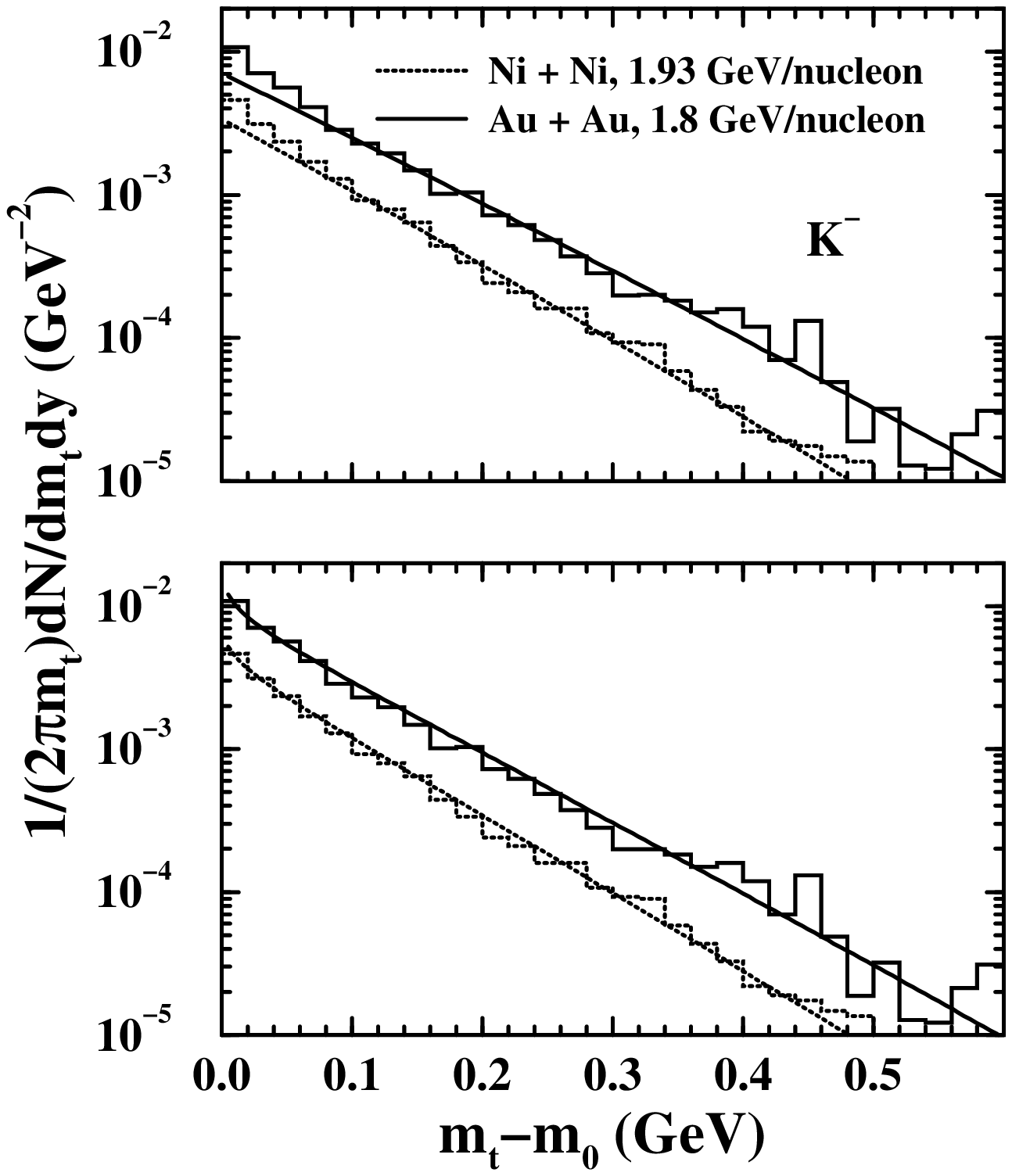}
\caption{
}
\label{fig4}
\end{figure}
\newpage
\begin{figure}
\leavevmode
\epsfxsize = 12cm
\epsffile[60 85 430 750]{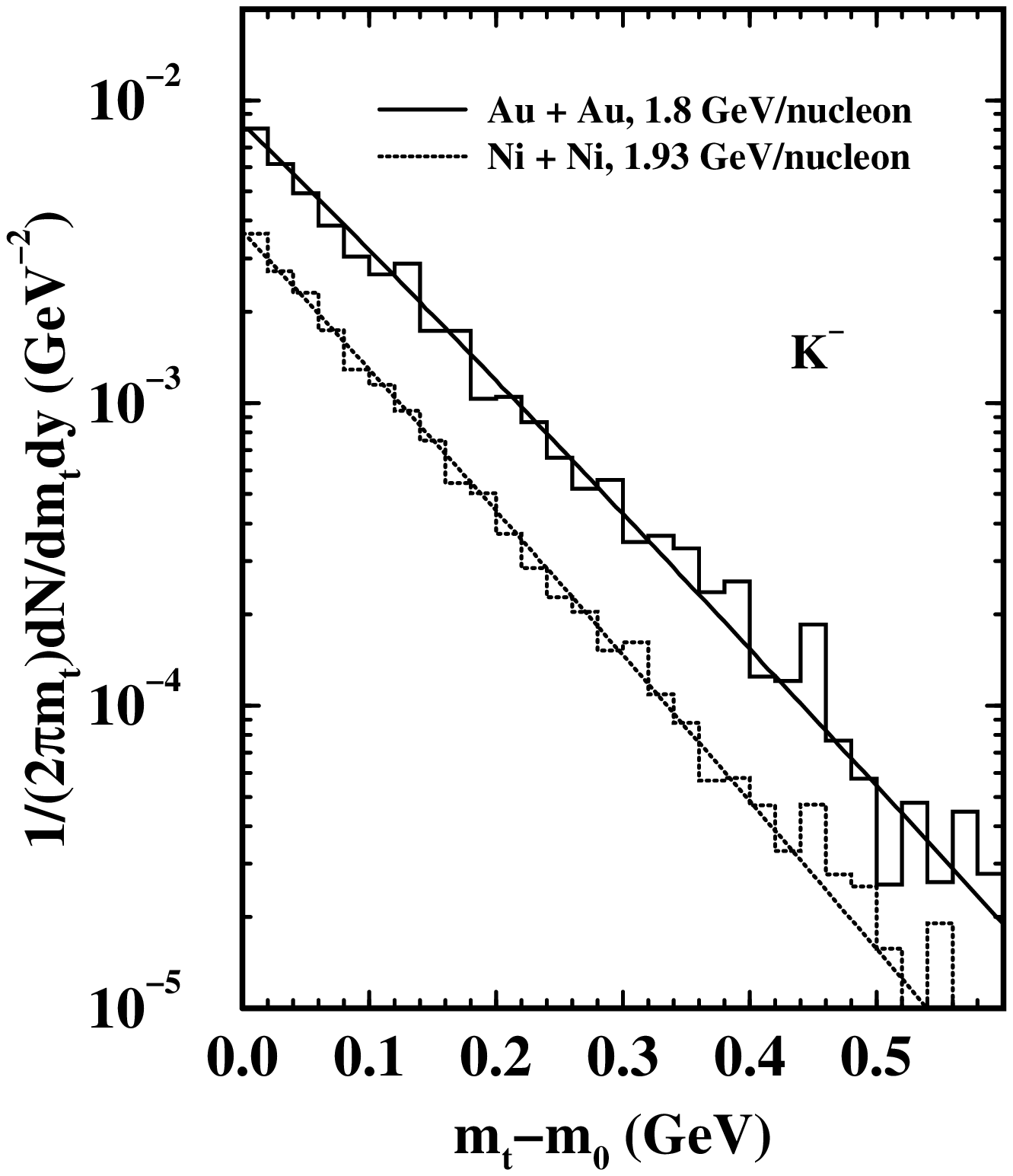}
\caption{
}
\label{fig5}
\end{figure}
\end{document}